\documentclass[aps,prl,twocolumn,showpacs]{revtex4}
\usepackage{graphicx}
\usepackage{amssymb}
\usepackage{color}
\begin{document}


\title{Wither the sliding Luttinger liquid phase in the planar pyrochlore}



\author{Marcelo Arlego and Wolfram Brenig}

\address{Institut f\"ur Theoretische Physik, Technische
Universit\"at Braunschweig, 38106 Braunschweig, Germany}

\email[E-mail: ]{m.arlego@tu-bs.de}


\date{\today}

\begin{abstract}
Using series expansion based on the flow equation method we study
the zero temperature properties of the spin$-1/2$ planar
pyrochlore  antiferromagnet in the limit of strong diagonal
coupling. Starting from the limit of decoupled crossed dimers we
analyze the evolution of the ground state energy and the
elementary triplet excitations in terms of two coupling constants
describing the inter dimer exchange. In the limit of weakly
coupled spin$-1/2$ chains we find that the fully frustrated inter
chain coupling is critical, forcing a dimer phase which
adiabatically connects to the state of isolated dimers. This
result is consistent with findings by  O. Starykh, A. Furusaki and
L. Balents (Phys. Rev. B \textbf{72}, 094416 (2005)) which is
inconsistent with a two-dimensional sliding Luttinger liquid phase
at finite inter chain coupling.
\end{abstract}

\pacs{ 75.10.Jm, 75.50.Ee, 75.40.$-$s, 78.30.$-$j}


\maketitle

\section{Introduction}

Geometrically frustrated quantum magnets above one dimension (1D)
have attracted strong interest in recent years \cite{diep-book}.
From an experimental point of view this is due in part to the
discovery of materials like the two-dimensional (2D) spin-dimer
system SrCu$_2$(BO$_3$)$_2$ \cite{2Dfm} and the three-dimensional
(3D) tetrahedral compounds Cu$_2$Te$_2$O$_5$X$_2$ with X=Cl, Br
\cite{3Dfm1, 3Dfm2} which exhibit strong frustration of the
magnetic exchange. From a theoretical point of view, the prime
interest in such systems stems from the possibility of realizing
quantum disordered states and spin liquid behavior. Particularly
promising candidates in this direction are systems which comprise
tetrahedra coupled into 2D or 3D networks. Indeed, experimental
observations on pyrochlore materials, like CsNiCrF$_6$
\cite{Harris94, Harris96} or Tb$_2$Ti$_2$O$_7$ \cite{Mirebeau04}
suggest the possibility of liquid-like behavior. In contrast to
the latter systems, which are essentially classical due to Ising
anisotropy and spins larger than 1/2, the $SU(2)$ invariant
spin-1/2 Heisenberg antiferromagnet on the 3D pyrochlore lattice
is an open issue. Experiments on the $S=1/2$ system
Y$_2$Ir$_2$O$_7$ \cite{Fukuzawa03} seem to indicate the presence
of quantum spin liquid state.
\begin{figure}[t]
\hspace{-0.75cm}\includegraphics[width=6.2cm]{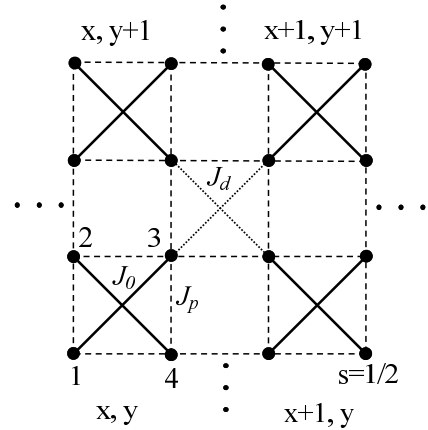}
\vspace{-2.5cm} \caption{The \emph{dimerized} planar pyrochlore
lattice. Spin$-1/2$ moments are located on solid circles. Crossed
dimers at site $x,y$ (bold solid lines) are locally coupled via
$J_p$ (dashed lines) and non-locally coupled to nearest neighbors
dimer by $J_p$ and $J_d$ (doted lines). On dimer exchange $J_0$
will be set to unity hereafter.} \label{cdm-lattice}
\end{figure}
To reduce the complexity of the 3D system, a 2D projection, i.e.
the planar pyrochlore, also known as checkerboard or crossed chain
model (CCM), which is shown in Fig.\ref{cdm-lattice} for
$J_p=J_d=J_0$, has been considered as a simplified starting point.
The CCM model and some of its anisotropic generalizations have
been studied by employing a variety of techniques. Analytic
methods include semiclassical $S \gg 1$ analysis \cite{Singh98,
Canals02, Tchernyshyov03}, large-N limit of the Sp(N) model
\cite{Chung01, Moessner04, Bernier04}, series expansion (SE) by
unitary transformation \cite{Brenig-Honecker02,
Brenig-Grzeschik04} and quantum field theory (QFT) methods
\cite{Starykh02, Starykh2005}. Besides, this model has been
studied extensively by means of exact diagonalization (ED)
techniques \cite{Palmer01,Fouet03, Sindzingre02}. These
investigations have given a fairly complete picture of the phases
of the system at different sectors of parameters space. At the
checkerboard point ($J_p=J_d=J_0$)\, a valence bond crystal (VBC)
ground state with a spin gap has been found \cite{Fouet03,
Brenig-Honecker02, Sindzingre02, Brenig-Grzeschik04}. For $J_d =
J_0 \rightarrow 0$ model of Fig.\ref{cdm-lattice} tends to the 2D
spin$-1/2$ Heisenberg model on the square lattice, which is known
to have gapless magnon excitations and AFM-LRO. From this limit,
by means of a phenomenological approach within the framework of
the Landau-Ginzburg-Wilson (LGW) theory, it has been suggested
recently  that the transition between the (AFM-LRO) phase and the
VBC, as $J_d ( = J_0)$ are increased, could happen within a
coexistence region or by a first order transition
\cite{Starykh2005} but this point deserves more investigation.
Previous results from ED up to 36 spins for $J_d = J_0$ in
ref.\cite{Sindzingre02} predict a finite gap at $J_d / J_p = 0.85$
and an extrapolated closure of the gap at $J_d / J_p = 0.65$.
Furthermore, semiclassical approaches predict stability of the
AFM-LRO phase, in the case of $S=1/2$ at $J_d = J_0$ for $J_d /
J_p \leq 0.76$.

In the opposite limit of weakly coupled spin$-1/2$ chains, i.e.
$J_p \ll J_d =J_0$ in Fig.\ref{cdm-lattice}, there are
controversial results. On the one hand, previous findings based on
QFT methods \cite{Starykh02} suggested that the model exhibits a
2D spin liquid ground state, being an example of the so called
\emph{sliding Luttinger liquid} (SLL) which is characterized by
the absence of LRO and the elementary excitations being massless,
deconfined spinons. It has been suggested that the prime reason
for this behavior stems from the momentum dependence of the
completely frustrated inter chain coupling. The latter vanishes
exactly at the AFM nesting vector, suppressing a conventional
magnetic instability. These results were corroborated subsequently
by numerical studies \cite{Sindzingre02}, where a range $J_p / J_d
\in [0, \simeq 0.8]$ for the extension of an effective 1D behavior
was suggested.

On the other hand, by employing bosonization and renormalization
group methods, Starykh et al. \cite{Starykh2005} have recently
suggested that instead of exhibiting a SLL phase, the model of
Fig.\ref{cdm-lattice} enters a gapped \emph{crossed dimer} phase
for $J_p \ll J_d =J_0$, with a spin gap $\Delta$ which opens
according to $\Delta / J_d \simeq 0.675 J_p^2$.

In this situation it is desirable to obtain results from
complementary methods, which will be the primary goal of this
work. Therefore, in the following analysis we will employ a SE in
terms of $J_p$ and $J_d$ as in Fig.\ref{cdm-lattice} and will
evaluate the ground state energy and the elementary excitations.
Our findings will support the existence of a crossed dimer phase,
in agreement with the results of ref.\cite{Starykh2005}.

To perform a SE, applicable in the vicinity of the anticipated SLL
phase we decompose the planar pyrochlore lattice into a set of
crossed, \emph{dimerized} chains as shown in
Fig.\ref{cdm-lattice}. I.e. the model represents a 2D array of AFM
coupled crossed spin$-1/2$ dimers. We will set $J_0=1$ hereafter.
The Hamiltonian reads
\begin{equation}\label{H}
  H = H_0 + V; \quad
  H_0 = \sum_{x,y} h_0; \quad
  V = \sum_{x,y} \left( V_p +  V_d \right),
\end{equation}
\noindent in which $h_0$ refers to the local crossed dimer unit at
site $x,y$, given by
\begin{eqnarray}
  h_0 &=& \left(\mathbf{S}_1 \cdot \mathbf{S}_3 +
  \mathbf{S}_2 \cdot \mathbf{S}_4 \right)  \label{H0} \\
   & &  + J_p ( \mathbf{S}_1 \cdot \mathbf{S}_2 + \mathbf{S}_2 \cdot \mathbf{S}_3
   + \mathbf{S}_3 \cdot \mathbf{S}_4 + \mathbf{S}_4 \cdot \mathbf{S}_1) \nonumber \\
   &=& \frac{1}{2}\left( \mathbf{S}_{13}^2 + \mathbf{S}_{24}^2 - 3 \right)
   + \frac{J_p}{2}\left( \mathbf{S}_{1234}^2 - \mathbf{S}_{13}^2 -
   \mathbf{S}_{24}^2
   \right),\nonumber
\end{eqnarray}
\noindent where all spins are at site $x,y$ and $\mathbf{S}_{1
\ldots n}=\mathbf{S}_1 + \ldots + \mathbf{S}_n$. \\
\noindent $V_p$ and $V_d$ in Eq.(\ref{H}) represent the couplings
between the crossed dimers at site $x,y$ with their next nearest
neighbors via $J_p$ and $J_d$ respectively, i.e.
\begin{eqnarray}
   V_p &=& J_p [ \mathbf{S}_{4; x,y} \cdot \mathbf{S}_{1; x+1,y}
   + \mathbf{S}_{3; x,y} \cdot  \mathbf{S}_{2; x+1,y} \\
&& + \mathbf{S}_{2; x,y} \cdot  \mathbf{S}_{1; x,y+1}
       + \mathbf{S}_{3; x,y} \cdot
       \mathbf{S}_{4; x,y+1}]; \nonumber \\
  V_d & = & J_d  [  \mathbf{S}_{3; x,y} \cdot  \mathbf{S}_{1; x+1,y+1}
  + \mathbf{S}_{2; x,y} \cdot  \mathbf{S}_{4; x-1,y+1} ].
  \nonumber \label{V}
\end{eqnarray}

\noindent Table \ref{tableH0} depicts the eigenstates of $h_0$.
\begin{table}[htb]
\begin{tabular}{c|c|c|c|c}
\hline   & $S_{1234}$ & $S_{13}$ & $S_{24}$ & $q=E + 3/2$
\\ \hline
$|{\cal S}_0\rangle$ & 0 & 0 & 0 & $0$ \\
$|{\cal T}_p\rangle$ & 1 & 1 & 0 & $1$ \\
$|{\cal T}_n\rangle$ & 1 & 0 & 1 & $1$ \\
$|{\cal S}_t\rangle$ & 0 & 1 & 1 & $2 - 2 J_p$ \\
$|{\cal T}_t\rangle$ & 1 & 1 & 1 & $2 - J_p$ \\
$|{\cal Q}_t\rangle$ & 2 & 1 & 1 & $2 + J_p$ \\
\hline
\end{tabular}
\caption{Eigenstates of a local crossed dimer unit $h_0$\,at site
$x,y$. Each state is labelled by the quantum numbers: $S_{1234}$,
$S_{13}$ and $S_{24}$ and the energy $E$.} \label{tableH0}
\end{table}
\noindent From Eq.(\ref{H0})\, it is clear that each state of
$h_0$ can be labelled by the following quantum numbers: total spin
$S_{1234}$, the spin of the dimer on the positive(negative)
diagonal, i.e. the $S_{13}(S_{24})$ and the energy $E$. As one can
observe, for $0 \leq J_p < 1$ the ground state is $|{\cal
S}_0\rangle$, i.e. both dimers are singlets. Besides, for $0 \leq
J_p < 1/2$ the first excited states are $|{\cal T}_p\rangle(|{\cal
T}_n\rangle)$ i.e. one of the dimers on the positive(negative)
diagonal is a triplet while the one on the other diagonal is a
singlet. The other states are: a total spin singlet $|{\cal
S}_t\rangle$, a triplet $|{\cal T}_t\rangle$ and a quintet $|{\cal
Q}_t\rangle$, all of them consisting of triplets on both
diagonals. Only the energies of the latter three states depend on
$J_p$.

\section{Series expansion by continuous unitary transformation}

In this section we briefly describe the SE expansion in terms of
$V$. To this end we rewrite Eq.(\ref{H}) as
\begin{equation}\label{Hrew}
    H=H_0(J_p=0) + J_p O_{p_0}^{l}+ \sum_{n=-N}^N \left( J_p O_{p_n}+
    J_d O_{d_n}\right).
\end{equation}
\noindent $H_0(J_p=0)$ has an equidistant ladder spectrum (Table
\ref{tableH0}) which we label with the \emph{total
particle-number} operator: $Q=\sum_{x,y}q(J_p=0)$. The
\emph{vacuum} corresponds to the $Q=0$ sector: $|0 \rangle \equiv
\prod_{x,y}|\mathcal{S}_{0; x,y}\rangle$. \emph{One-particle}
states correspond to the $Q=1$ sector, whose basis states
$|T_{\mu; x,y}\rangle (\mu={p,n})$ are composed of a local triplet
on the positive or negative diagonal excited
 from the vacuum: $|T_{\mu ; x', y'}\rangle \equiv
|\mathcal{T}_{\mu; x', y'}\rangle \bigotimes \prod_{x,y \neq
x',y'}|\mathcal{S}_0; x,y \rangle$. The $Q \geq 2$ sector is of
multi-particle nature.

The $O_{p_0}^{l}$ operators in Eq.(\ref{Hrew}) refer to the local
contributions proportional to $J_p$ in $H_0$. Finally, the third
term in the same Eq. represents the inter dimer coupling via $V$.
The $O_{p_n}$ and $O_{d_n}$ operators \cite{Oelem} non-locally
create(destroy) $n > 0 (<0)$ energy quanta within the ladder
spectrum of $H_0(J_p=0)$ . Note that only terms containing two
dimers at different $x,y$ sites are present in $V$. This imposes
the constraint $N \leq 2 $.

Models of type Eq.(\ref{Hrew}) allow for SE using the flow
equation method of Wegner \cite{Wegner94,Knetter00}. There, by
means of a continuous unitary transformation, one maps $H
\rightarrow H_{eff}$ with
\begin{equation}\label{Heff}
    H_{eff} = H_0 + \sum_{r, 0\leq s \leq
    r}C_{r,s}J_p^{r-s} J_d^s,
\end{equation}
\noindent in which the $C_{r,s}$ operators involve  $r-s$ and $s$
products of type-$O_{p_n}$ and  $O_{d_n}$ operators, respectively.
We refer to \cite{Knetter00} for further details.

The main advantage of the flow equation method is that unlike $H$,
the effective Hamiltonian $H_{eff}$ is constructed to
\emph{conserve} the total number of particles $Q$ of $H_0(J_p=0)$
at each order in the expansion. This allows for the SE of several
quantities from the bare eigenstates of $H_0(J_p=0)$ as we show in
the following sections.

\section{Ground state energy}
The ground state energy allows for an important consistency check
of our calculation with respect to existing literature and other
SE results. As mentioned above, there is evidence for a {\em
quadrumer} VBC phase of the CCM at the checkerboard point
($J_p=J_d=1$). This is based on ED \cite{Fouet03} and SE starting
from the quadrumer limit of the CCM
\cite{Brenig-Honecker02,Brenig-Grzeschik04}. Moreover, Starykh et
al. \cite{Starykh2005}, have predicted a first order transition or
a coexistence phase at some intermediate value of $0 < J_p < 1$
and $J_d =1$ between the VBC and a crossed dimer phase. Therefore,
in order for the present SE to give results consistent with them
for $J_p$ and $J_d$ in the vicinity of the checkerboard point, we
expect the ground state energy of the crossed dimer state to be
larger than that of the quadrumer VBC at $J_p=J_d=1$.

For $0 \leq J_p < 1$ the ground state of $H_0(J_p=0)$ is
the $Q=0$ sector, i.e. $|0 \rangle \equiv
\prod_{x,y}|\mathcal{S}_0\rangle_{x,y}$. Q-conservation of
$H_{eff}$ allows for the calculation of the evolution of this
state as a SE($J_p, J_d$) by means of
\begin{equation}\label{Egccm}
    E_g = \langle 0 | H_{eff} | 0 \rangle.
\end{equation}
\noindent The actual calculation can only be done on finite
clusters. To obtain results to $O(J_p^n, J_d^n)$, valid in the
thermodynamic limit, i.e. for infinite sized systems, matrix
elements given by Eq.(\ref{Egccm}) have to be evaluated on
clusters with periodic boundary conditions (PBC), sufficiently
large to avoid a wrap-around at graph length $n$ \cite{step-note}.
In our case, the size of the Hilbert space, considering both
perturbations $J_p$ and $J_d$, restricts the calculation at $n=5$
due to memory limitations. We find this enough to quantitatively
describe the combined effects of $J_p$ and $J_d$ in the range of
interest. At $O(J_p^5, J_d^5)$ the ground state energy per spin
and PBC, $e_g(J_p, J_d)$, is given by
\begin{eqnarray}\label{egccm}
 & & e_g (J_p, J_d) = -\frac{3}{8}-\frac{3}{64}J_d^2 -\frac{3}{32}J_p^2
  -\frac{3}{256}J_d^3 + \frac{9}{128}J_d J_p^2 \nonumber \\
   & & -\frac{3}{128}J_p^3 -\frac{13}{4096}J_d^4
   -\frac{7}{1024}J_d^2 J_p^2 + \frac{5}{128}J_d J_p^3 -\frac{47}{2048}J_p^4 \nonumber \\
   & &  -\frac{89}{49152}J_d^5
   -\frac{307}{24576}J_d^3 J_p^2 -\frac{71}{12288}J_d^2 J_p^3
   \nonumber \\
   & &  + \frac{869}{24576} J_d J_p^4 -\frac{175}{8192}J_p^5.
\end{eqnarray}
As a check, note that $e_g (J_p=0, J_d)$ reduces to the $O(J_d^5)$
dimer expansion of the ground state energy  of the dimerized
spin$-1/2$ chain \cite{Knetter00, Barnes99}.

From Bethe ansatz \cite{Bethe31} the ground state energy for the
homogeneous spin$-1/2$ chain is known: $\widetilde{e_g} (J_p=0,
J_d=1)= 1/4 - \ln 2$ \cite{Hulthen38}. Comparing it with $e_g
(J_p=0, J_d=1)$ from Eq.(\ref{egccm}) we obtain a relative error
of $\approx 1 \%$, which gives a rough estimation of the
convergence \cite{conv-note}.

\begin{figure}[t]
\vspace{-0.7cm}
\includegraphics[angle=-90, width=0.48\textwidth]{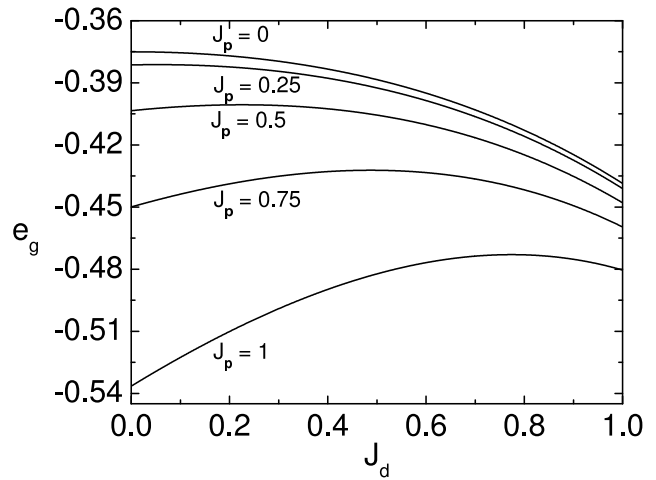}
\caption{Ground state energy $e_g (J_p, J_d)$ per spin vs $J_d$
for different values of $J_p$, Eq.(\ref{egccm}). As observed
$e_g(J_p, J_d)$ monotonously decreases  as $J_p$ is increased from
$0$ to $1$, for fixed values of $J_d$. Eq.(\ref{egccm}) at the
point of decoupled uniform spin$-1/2$ chains, i.e.  $e_g (J_p=0,
J_d=1)$, differs $\approx 1 \%$ from the exact value $1/4 - \ln
2$\, from Bethe ansatz. At the checkerboard point Eq.(\ref{egccm})
gives $e_g (J_p=1, J_d=1) \approx -0.48$ which is above the value
obtained by employing a SE starting from a different point in
parameters space (see text). This is consistent with the stability
of the quadrumer VBC in the vicinity of the checkerboard point
which has $\overline{e}_{g} \approx -0.51$.} \label{eg-plot}
\end{figure}

Fig.\ref{eg-plot} depicts the evolution of the ground state energy
as given by Eq.(\ref{egccm}) vs $J_d$ for different values of
$J_p$. As shown, $e_g(J_p, J_d)$ monotonously decreases as $J_p$
is incremented from $0$ to $1$, for fixed values of $J_d$. The
position $J_{d_{m}}$ of the maximum of $e_g(J_p, J_d)$ vs $J_d$
shifts from $J_{d_{m}}= 0$ to $J_{d_{m}}\approx 0.8$ for $J_p$
taking fixed values from $0$ to $1$, respectively.
At the checkerboard point our SE yields $e_g(J_p=1,J_d=1)\approx
-0.48$. Note that our value is larger than the value of
$\overline{e}_{g} \approx -0.51$ which has been obtained from the
quadrumer SE in ref.\cite{Brenig-Honecker02}. While the expansion
parameters and their respective orders cannot be compared directly
between the present SE and that of ref.\cite{Brenig-Honecker02}
these findings are at least consistent with the stability of the
quadrumer VBC in the vicinity of the checkerboard point. This,
together with the fact that $H_0(J_p=1)$ has a degenerated ground
state (see Table 1), suggests that Eq.(\ref{egccm}),\, at least at
the checkerboard point, does not represent the ground state energy
of the system anymore, which is $\overline{e}_{g} \approx -0.51$.

\section{Triplet dispersion}
Next we evaluate the dispersion of the one-triplet states. To this
end it is necessary to diagonalize $H_{eff}$ in the $Q=1$ sector
of $H_0(J_p)$, i.e. the subspace spanned by the vectors $|T_{\mu;
x,y}\rangle, \mu=\{p,n\}$. First we switch to momentum space
\begin{equation}\label{fourier}
|T_{\mu;k_x,k_y}\rangle = \sqrt{\frac{1}{L^2}}
    \sum_{x,y}\left[\exp  \left(i x k_x  + i y k_y \right)
    |T_{\mu ; x,y}\rangle\right],
\end{equation}
in which PBC is assumed and $L^2$ is the number of sites. Because
of the degeneracy of the Q=1 sector, $H_{eff}$ is a $2 \times 2$
matrix in the basis given by Eq.(\ref{fourier}), whose entries $
E_{T \mu, \nu ; k_x, k_y}$ are
\begin{eqnarray}\label{matrix-trip}
  E_{T \mu, \nu ; k_x, k_y} &=&  \langle T_{\mu ; k_x, k_y}| H_{eff} | T_{\nu ; k_x, k_y}\rangle
  \nonumber  \\
   &=& \sum_{x,y}\left[ c_{\mu, \nu ; x,y}\exp \left( i x k_x + i y k_y \right) \right],
\end{eqnarray}
where $\mu, \nu = \{p,n\}$ and
\begin{equation}\label{hoppings}
    c_{\mu, \nu ;  x,y}  =   \langle T_{\mu ; x,y} | H_{eff} |
    T_{\nu ; 0,0}
    \rangle
\end{equation}
\noindent are the hopping amplitudes from a $\nu-$type of triplet
on site $0,0$ to a $\mu-$type of triplet on site $x,y$. Evaluation
of the $ c_{\mu, \nu ; x,y}$'s is facilitated by symmetry which
leaves only a subset of them to be independent. To obtain the
hopping amplitudes valid in the thermodynamic limit, the $ c_{\mu,
\nu ; x,y}$'s are calculated on finite clusters with open boundary
conditions (OBC), and a cluster-topology suitable to embed all the
paths of length $n$ connecting the site $(0,0)$ with $(x,y)$ at
$O(n)$ of the perturbation. From this, the triplet dispersion is
obtained as
\begin{equation}\label{disp-eq}
    \omega(\textbf{k}, J_p, J_d) = E_T (\textbf{k}, J_p, J_d) - \langle 0
    |H_{eff} | 0 \rangle ,
\end{equation}
where $\textbf{k}=(k_x, k_y)$ and the second term refers to the
ground state energy which has to be evaluated on the identical
cluster as $ c_{\mu, \nu ; 0,0}$. The triplet energy $E_T
(\textbf{k}, J_p, J_d)$ results from diagonalizing the $2 \times
2$ matrix of Eq.(\ref{matrix-trip}), leading to two excitation
branches.

We have calculated the triplet dispersion $\omega(\textbf{k}, J_p,
J_d)$ up to $O(J_p^5, J_d^5)$ \cite{disp-expl}. Upon rotating
$k=(k_x, k_y) \rightarrow (k_p, k_n)$ along the positive and
negative diagonals, by means of $k_{x,y}= (k_p \mp k_n)/2$, the
two branches of Eq.(\ref{disp-eq}) have to reduce to the triplet
dispersion of the dimerized spin$-1/2$ chains along positive and
negative diagonals for $J_p=0$. For consistency, we have checked
that this is satisfied up to the order of our SE by comparing with
existing SE results on the single dimerized chain \cite{Knetter00,
Barnes99}.

\section{Triplet gap}
Using the results from the previous section we may now turn to one
of the main objectives of this work, i.e. to study the evolution
of the triplet gap under the combined effect of $J_p$ and $J_d$.
This will enable us to conclude on the stability of the
dimerization along the chains and therefore to discriminate
between the controversial proposals on the stability of the SLL.
The triplet gap: $\Delta (J_p, J_d) = \textrm{min}\,
\omega(\textbf{k}, J_p, J_d)$ is obtained from the minimum in $\bf
k$-space of the two one-triplet branches. For $J_p \leq 0.5$ we
find the triplet gap to be at $k_p = k_n = 0$. It is given by
\begin{eqnarray}\label {ccm-gap}
  & & \Delta(J_p, J_d) = 1 - \frac{1}{2} J_d  -\frac{3}{8} J_d^2
  +\frac{1}{32} J_d^3 + \frac{29}{32}J_d J_p^2 -\frac{3}{32} J_p^3  \nonumber \\
  & &   -\frac{5}{384} J_d^4 -\frac{401}{384} J_d^2 J_p^2
   + \frac{17}{32} J_d J_p^3 -\frac{53}{128}J_p^4 -\frac{761}{12288}J_d^5 \nonumber  \\
  & & + \frac{1309}{1536}J_d^3 J_p^2
  -\frac{3895}{4608} J_d^2 J_p^3 + \frac{6253}{3072} J_d J_p^4 - \frac{2065}{4608}
  J_p^5.
 \end{eqnarray}

We have analyzed $\Delta(J_p, J_d)$ both by its plain SE and by
using a biased-Pad\'e approximation\cite{Knetter00}. The latter
consists of an extrapolation biased by the known analytical form
of the gap-closure of the dimerized spin$-1/2$ chain as a function
of $J_d$ \cite{Orignac2004}, i.e. $\Delta(J_p=0, J_d \rightarrow
1) \propto (1 - J_d) ^{2/3}$. Using this, $J_d$ is replaced in
favor of $J_d = 1 - x^{3/2}$ in Eq.(\ref{ccm-gap}).  Then,
standard Pad\'e approximation\cite{Guttmann89} is performed around
$x=1$ and finally $x=(1 - J_d)^{2/3}$ is re-inserted, arriving at
a transformed Pad\'e approximation in terms of  the original
variable $J_d$.

\begin{figure}[t]
\vspace{-0.3cm}\hspace{-0.6cm}\includegraphics[angle=-90,
width=0.49\textwidth]{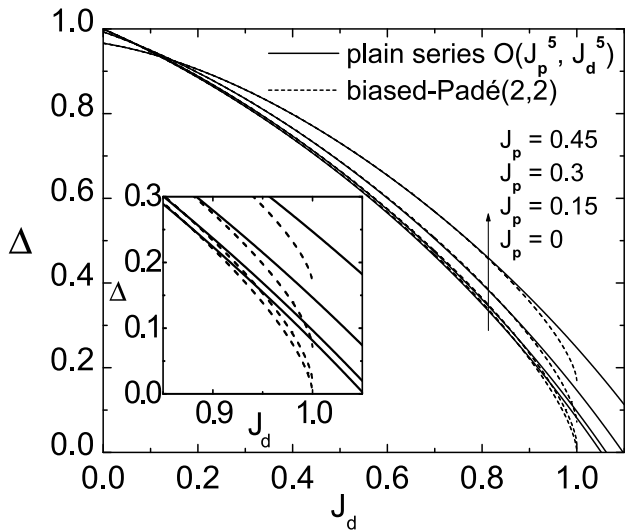}
\caption{Triplet gap $\Delta$
vs $J_d$ for increasing values of $J_p$, Eq.(\ref{ccm-gap}) (solid
lines) and biased-Pad\'e(2,2) (dashed lines). As the inset shows,
the biased-Pad\'e(2,2) describes the closure of the gap at ($J_p=0
, J_d=1)$ very well. The effect of $J_p$ opening the gap is
evident from the arrow.} \label{gap-Jd}
\end{figure}

Fig.\ref{gap-Jd} depicts the evolution of $\Delta(J_p, J_d)$ as a
function of $J_d$ for different values of $J_p$. Solid lines show
the results from the plain series, i.e. Eq.(\ref{ccm-gap}) and
dashed lines exhibit the biased-Pad\'e(2,2) approximants. As the
inset of Fig.\ref{gap-Jd} shows, the approximant describes the
closure of the gap at $J_p=0, J_d=1$ very well. This justifies the
use of this extrapolation technique and shows that our expansion
can be employed to analyze the coupling of the spin$-1/2$ chains
via $J_p$.

As a first result from Fig.\ref{gap-Jd}, both the plain series and
the biased-Pad\'e clearly show an \emph{opening} of the gap at
$J_d =1$ for increasing values of $J_p$. As it is also obvious
from these plots, this tendency persists for all $J_d \geq 0.2$.
Second, Fig.\ref{gap-Jd} shows, that the biased Pad\'e approximant
leads to observable differences from the plain SE only in the
vicinity of the gap-closure.
\begin{figure}[t]
\begin{center}
\vspace{-0.4cm} \hspace{-0.62cm}\includegraphics[angle=-90,
width=0.515\textwidth]{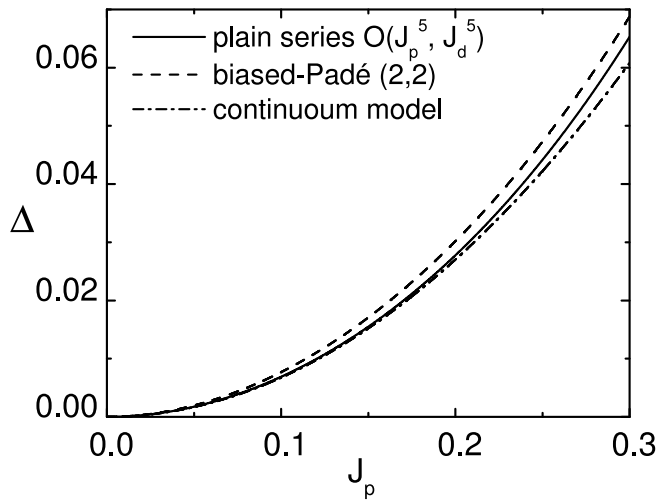}
\end{center}
\caption{Evolution of the triplet gap $\Delta(J_p, J_d=1) -
\Delta(J_p=0, J_d=1)$ vs $J_p$ starting from the limit of
decoupled spin$-1/2$ chains ($J_p=0, J_d=1)$. Plain series,
Eq.(\ref{ccm-gap}), (solid line) and biased-Pad\'e(2,2)
approximant (dashed line). The shift $\Delta(J_p=0, J_d=1)$ has no
effect on the functional dependence on $J_p$ and it is made for a
better comparison with the QFT results by Starykh et al.
\cite{Starykh2005}, i.e. $\Delta (J_p \ll 1, J_d=1)\approx 0.675
J_p^2$, (dash-doted line). Very good agreement between both
techniques, up to $J_p \approx 0.1$, is observable. The same
agreement is found by employing standard Dlog-Pad\'e technique.}
\label{gap-Jp}
\end{figure}

Now, we consider the opening of the spin gap at $J_d=1$ for $J_p
\ll 1$. To this end, we first shift the zero of energy in order to
suppress the finite gap of the SE at $J_p$=0 and $J_d$=1, i.e. we
consider $\Delta(J_p, J_d=1) - \Delta(J_p=0, J_d=1)$. By this we
account for the inability of the SE to provide for a truly gapless
spectrum in the limit of decoupled AFM Heisenberg chains. As it is
obvious from Fig.\ref{gap-Jd}, this shift is very small for the
biased Pad\'e, i.e. $\approx 0.003$ and $\approx 0.081$ for the
plain SE. The resulting spin gap versus $J_p$ is shown in
Fig.\ref{gap-Jp} where the solid line represents the plain series
and the dashed line corresponds to the biased-Pad\'e(2,2)
approximant.

These results may be compared with recent findings of Starykh et
al. \cite{Starykh2005}. Using QFT methods, i.e. renormalization
group and bosonization, they have obtained an analytic expression
for the spin gap, which is quadratic in $J_p$, namely $\Delta (J_p
\ll 1, J_d=1) = c J_p^2$ where $c \approx 0.675$. This function is
shown in the same Fig.\ref{gap-Jp} with a dash-dotted line.
Obviously, there is a very good agreement up to $J_p \approx 0.1$.
This is remarkable, given the completely different nature of our
approach and that of ref.\cite{Starykh2005}. We note that the same
agreement is found when we employ standard Dlog-Pad\'e
approximants instead of biased-Pad\'e. Therefore, we are strongly
tempted to claim that the stable phase of the CCM at finite $J_p$
and at $J_d=1$ is indeed gapped. This is inconsistent with a SLL
state.

While Fig.\ref{gap-Jp} seems to suggest an ever-increasing spin
gap, one has to remember, that at $J_p = 0.5 $ the nature of the
first excited state of $h_0(J_p)$ changes from the triplet $|
\mathcal{T}_{p,n}\rangle$ to the singlet $| \mathcal{S}_t \rangle
$ (Table 1). This imposes an upper limit on the range of validity
of the SE.

\section{Conclusions}

In summary, using series expansion, we have studied the zero
temperature properties of the planar pyrochlore quantum magnet
focussing on its behavior close to the crossed chain limit. In
particular, we have considered the ground state energy, the
elementary triplet dispersion and the spin gap.

In relation with the ground state energy calculations our findings
are consistent with previous results regarding the stability of
the quadrumer VBC at the checkerboard point.

Our main result regards the existence of crossed dimer phase. Our
calculations univocally show an opening of the gap $\Delta(J_p \ll
1, J_d = 1)$. Moreover, our expansion (Eq.(\ref{ccm-gap})), allows
to route the physical origin of the crossed dimer phase by showing
that it is adiabatically connected with decoupled crossed dimer
units.

On the other hand, the very good agreement with a recent field
theory approach by Starykh et.al. \cite{Starykh2005} even for $J_p
\leq 0.1$ consolidates the hypothesis about the existence of a
crossed dimer phase in the region of parameters space $J_d=1$,
$J_p \ll 1$ and indicates the absence of Luttinger liquid
behavior, the so called sliding Luttinger liquid phase, in this
region of parameters space.

\section{Acknowledgments}
The authors thank A. Honecker for useful discussions. This
research was supported in part through DFG Grant No. BR 1084/4-1.


\begin{thebibliography}{99}
\bibitem{Oelem} Explicit tabulation of $O_{p_n}$ and $O_{d_n}$
is electronically available on request from the authors.
\bibitem{step-note} A graph of lenght $n$ is composed by $n$ steps.
Each step connects a given site with its nearest neighbors along
both $x,y$ directions and the diagonals.
\bibitem{conv-note}The convergence of the SE for $e_g$ may be asymptotic. The
evaluation of $e_g(J_p=0,J_d^{10})$ \cite{Knetter00} only reduces
the error $\approx 0.4 \%$).
\end{thebibliography}
\end{document}